\documentclass[%
 aip,
 amsmath,amssymb,
 reprint,%
]{revtex4-1}

\usepackage{graphicx}% Include figure files
\usepackage{dcolumn}% Align table columns on decimal point
\usepackage{bm}% bold math
\usepackage[english]{babel}
\usepackage[utf8]{inputenc}
\usepackage[T1]{fontenc}
\usepackage{mathptmx}
\usepackage{etoolbox}
\usepackage{xcolor}
\makeatletter
\def\@email#1#2{%
 \endgroup
 \patchcmd{\titleblock@produce}
  {\frontmatter@RRAPformat}
  {\frontmatter@RRAPformat{\produce@RRAP{*#1\href{mailto:#2}{#2}}}\frontmatter@RRAPformat}
  {}{}
}%
\makeatother
\begin{document}

\preprint{AIP/123-QED}

\title[The Chaotic Milling Behaviors of Interacting Swarms After Collision]
{The Chaotic Milling Behaviors of Interacting Swarms After Collision}
% Force line breaks with \\
\author{Sayomi Kamimoto}

\altaffiliation[SK is a ]{National Research Council Postdoctoral Research Associate.}
\email{sayomi.kamimoto.ctr.ja@nrl.navy.mil, jason.hindes@nrl.navy.mil, ira.schwartz@nrl.navy.mil}

\author{Jason Hindes}%
% \email{jason.hindes.ctr@nrl.navy.mil}
\affiliation{ U.S. Naval Research Laboratory, Washington, DC 20375, USA.}
%Authors' institution and/or address%\\This line break forced with \textbackslash\textbackslash
%}%

\author{Ira B. Schwartz}
\affiliation{ U.S. Naval Research Laboratory, Washington, DC 20375, USA.}
%Second institution and/or address%\\This line break forced% with \\}%

\date{\today}% It is always \today, today,

\begin{abstract}
We consider the problem of characterizing the dynamics of interacting swarms
after they collide and form a stationary center of mass.  Modeling efforts have shown that the collision of near head-on interacting swarms can produce a variety of post-collision dynamics including coherent milling, coherent flocking, and scattering behaviors.  In particular,  recent analysis of the transient dynamics of two colliding swarms has revealed the existence of a critical transition whereby the collision results in a combined milling state about a stationary center of mass. In the present work we show that the collision dynamics of two swarms that form a milling state transitions from periodic to chaotic motion as a function of the repulsive force strength and its length scale.  We used two existing methods as well as one new technique: Karhunen-Loeve decomposition to show the effective modal dimension chaos lives in,   the 0-1 test to identify chaos,  and then Constrained Correlation Embedding to show how each swarm is embedded in the other when both
swarms combine to form a single milling state after collision.  We expect our analysis
to impact new swarm experiments which examine the interaction of multiple swarms.
\end{abstract}
\maketitle
\begin{quotation}
%Swarms, which consist of agents with simple rules are ubiquitous in nature and exhibit many different types of emergent dynamical behavior.  Much recent work has explored the dynamics of single-swarm pattern formation, while much less is known about the behaviors of multiple mutually interacting and colliding swarms.  Applying existing methods of analyzing the dynamics of colliding swarms,  we show the onset of deterministic chaos as a function of repulsion strength. 
Swarms, which consist of agents with simple rules,  are ubiquitous
in nature and exhibit many different types of emergent dynamical behavior.
Much recent work has explored the dynamics of single swarm that depend on
general physical parameters, such as attraction, repulsion, alignment
and communication delay.  The collective patterns have been derived
from bifurcation theory and mean-field analyses.  However,  much less
is known about the dynamical behavior of multiple interacting and colliding 
swarms.  Colliding swarms have been found to
have collective behaviors, such as milling,  flocking and scattering states.  Milling,  in particular, is created
when one swarm captures another resulting in a stationary center of
mass.   Applying existing
methods of analyzing the dynamics of colliding swarms,  we describe the long term dynamics of the captured milling states.  Specifically,  we show the resulting dynamics
starts from periodic behavior,  going through quasi-periodic
behavior on a torus,  which then leads to chaotic behavior as a function of repulsive strength.  
%We note that
%accompanying such a transition, the modal dimension of the attractors
%goes through a phase transition that occurs at a critical set of parameters.
%In addition to the high dimensional chaotic behavior of the captured
%milling state,  we describe the mixing of the two swarms by measuring
%local complexity of one swarm with respect to another. 
\end{quotation}
\section{\label{sec:Intro}Introduction}
Improved data gathering and analysis techniques have enabled
researchers to collect data and analyze the motions of individual agents in
biological flocks,  and formulate more accurate, empirical models for
collective motion strategies of flocking species such as birds, fish and
insects,  \cite{Li_Sayed_2012,Theraulaz2002}.  The derived mechanics has
  resulted in the translation of swarm theory to communicating robotic
systems.  Swarms of collaborating robots have been proposed for conducting tasks
including search and rescue, density control, and mapping \cite{Li17,Ramachandran2018}.

From general models of swarm dynamics, one observes three basic swarming states or modes: flocking, in which a center of mass moves in a straight line with a constant velocity; milling, where the agents rotate coherently around a stationary center of mass; and rotating, where the center of mass itself rotates about a fixed point in space \cite{Szwaykowska2018,
edwards2020delay,CollidingSwarms2021}.  Much is known about the behaviors and stability of single swarms with physics-informed, nonlinear interactions \cite{Levine,Minguzzi,DOrsagna,Romero2012}.  They are able to converge to organized, coherent behaviors in spite of complicating factors such as communication delay,
localized number of neighbors each agent is able to interact with,  a consensus force,
heterogeneity in agent dynamics, and environmental noise \cite{Ouellette19, Mier, Szwaykowska2016, schwartz2020torus}.  However,  in many applications,  beyond the dynamics of single swarms,  multiple swarms interact and produce even more complex spatio-temporal behaviors. 

New studies have begun to address swarm-on-swarm dynamics,  and in particular the scattering of two large, colliding swarms with nonlinear interactions.  Recent numerical studies have shown that when two flocking swarms collide, the resulting dynamics typically appear as a merging of the swarms into a single flock, milling as one uniform swarm, or scattering into separate composite flocks moving in different directions \cite{Armbruster2017Elastic, kolon2018dynamics, Sartoretti}.  However, more detailed analytical understanding of how and when
these behaviors occur is necessary,  especially when designing robotic swarm experiments such as swarm herding and capture \cite{9029573,9303837,8444217} and controlling their outcomes. 

\begin{figure}
\begin{center}
\includegraphics[width=9.5cm]{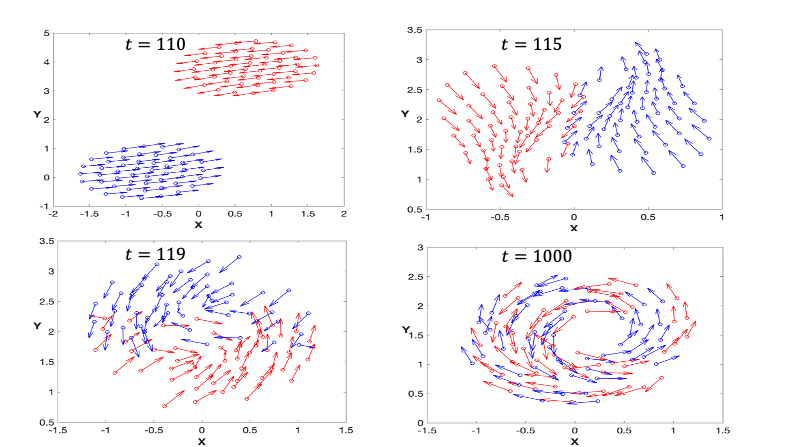}
\caption{\label{fig:SNAP} Four time-snapshots for different values of t,  showing each swarm with different
colors: red and blue circles. Velocities are drawn with arrows.  Swarm parameters are: C = 1.15,  N = 50. }
\end{center}
\end{figure}

Building on early numerical insights,  it has been shown that one can 
predict the critical swarm-on-swarm interaction coupling, below which two colliding swarms
merely scatter, as a function of a physical swarm parameter
\cite{CollidingSwarms2021}.  In particular,  analysis of the transient dynamics of two
colliding swarms,  seen in Fig. \ref{fig:SNAP},  has revealed the existence of a critical transition whereby
the collision results in a combined milling state,  where one swarm is
  embedded in the other.  However, what is missing is a quantitative
  description of the post-collision asymptotic dynamics.
  
\begin{figure*}
\begin{center}
\includegraphics[width=5.925cm]{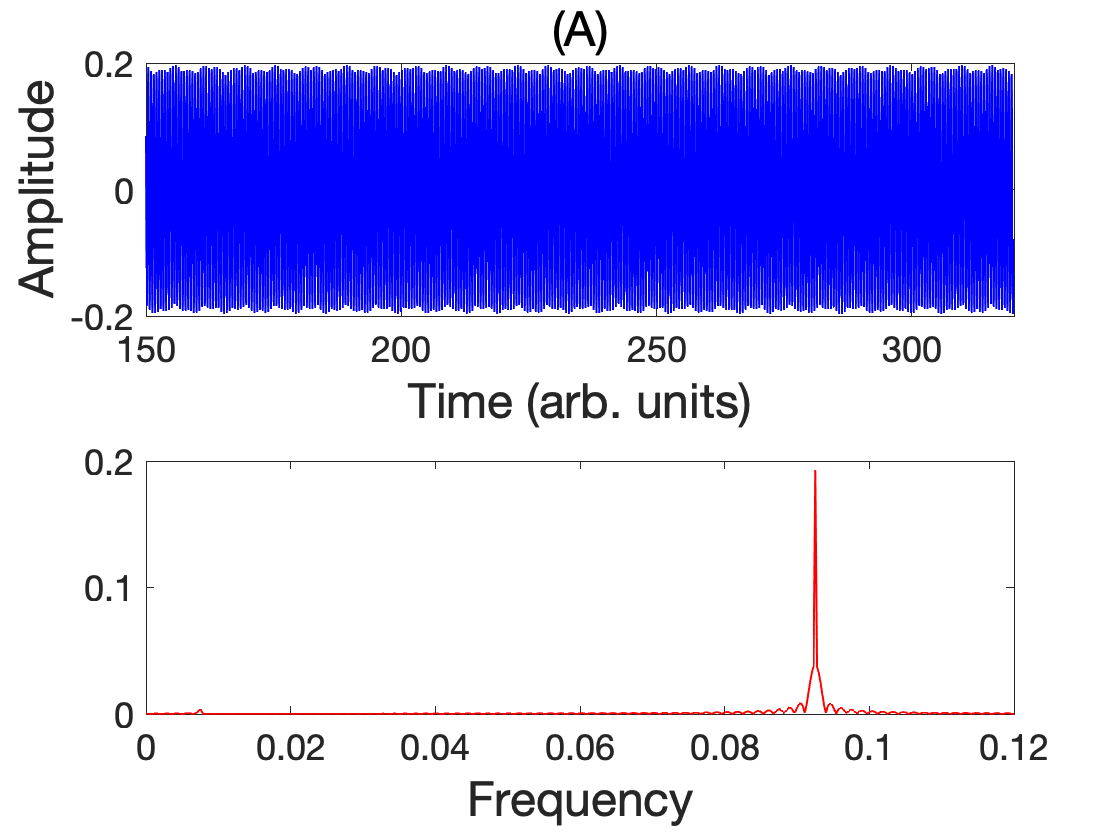}
\includegraphics[width=5.925cm]{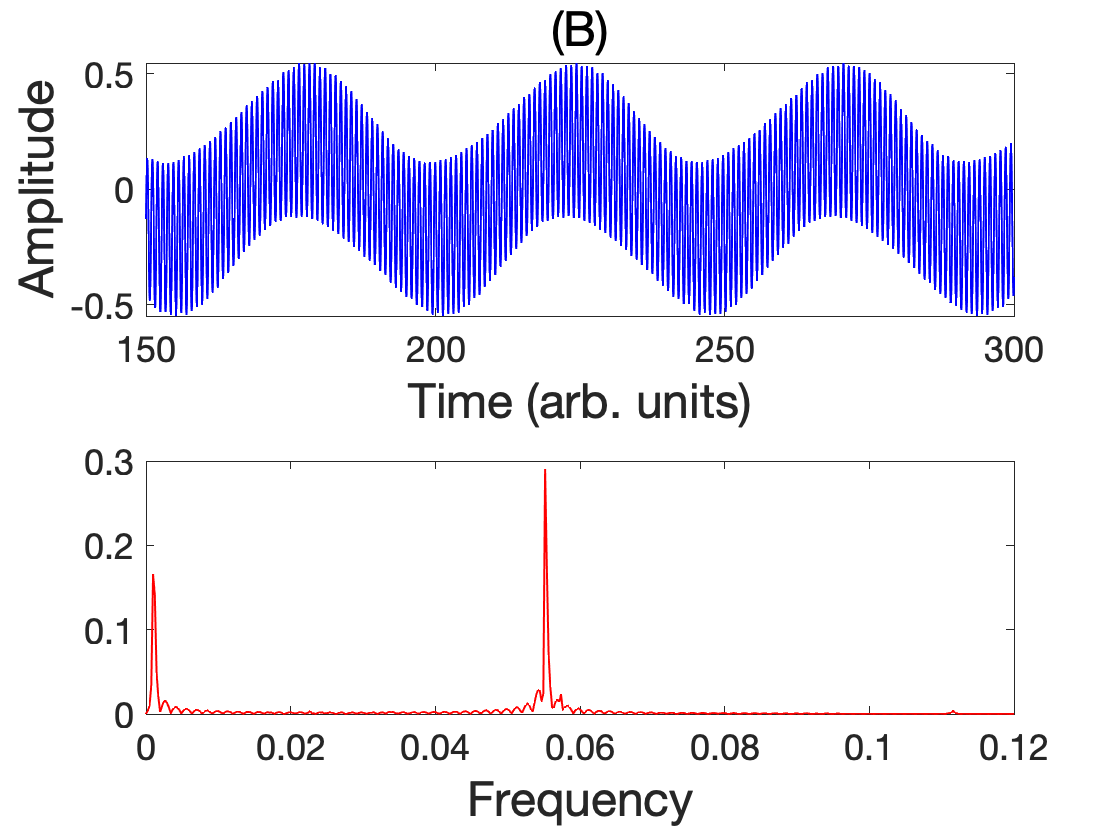}
\includegraphics[width=5.925cm]{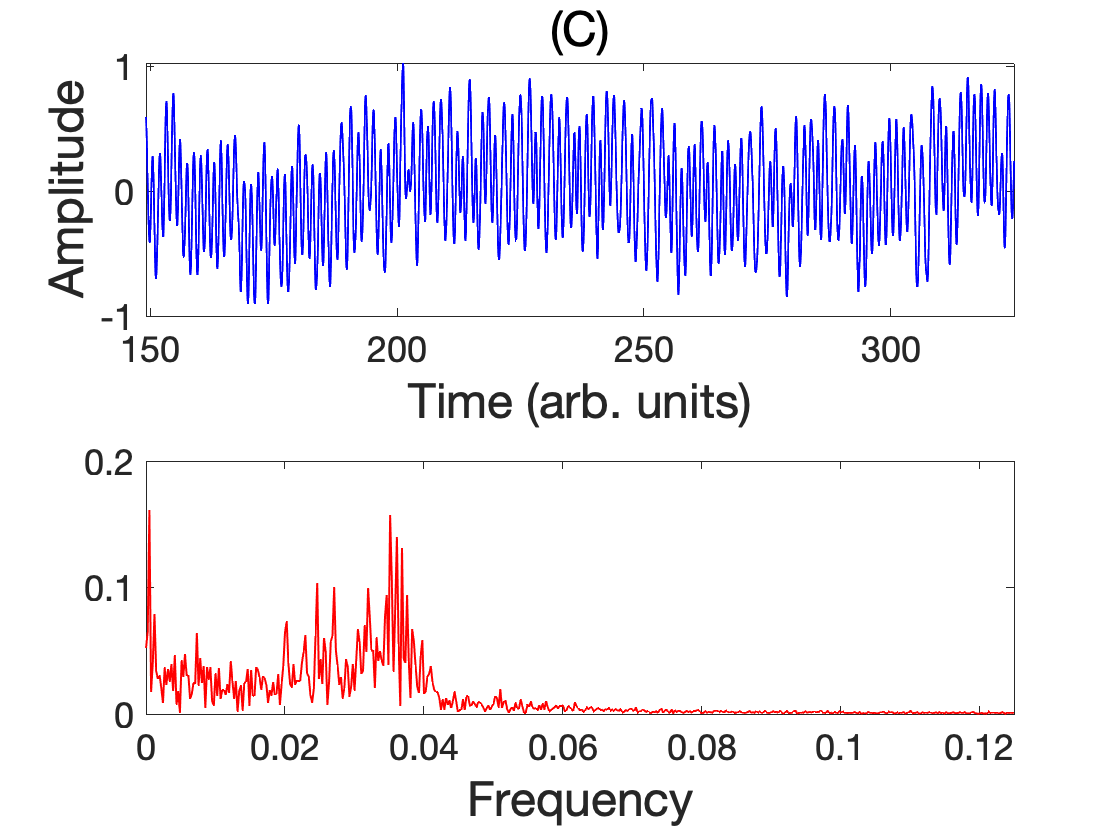}
\end{center}
\caption{\label{fig:TS} Dimensionless time-series (top) and frequency (bottom) of y component of a single agent randomly chosen,  for (A): C = 0.51,  (B): C = 0.70  and (C):C = 0.89.  Accordingly, we have observed periodicity (A),  a motion on torus (B) and chaos (C). }
\end{figure*}

In the current work,  our
starting point is Fig.~\ref{fig:TS}. We observe that after the colliding swarms have settled down to a milling behaviors,  the dynamics change from periodic to quasi-periodic,  and to chaotic motions for different values of repulsion strength.  We examine the dimensionality of the combined mill by computing its Karhunen-Loeve dimension. It is a measure of the variance of the entire post-collision attractor, and computes how many expansion modes it takes to capture the dynamics.  Next, we aim to numerically confirm that
the two swarms near collisions displayed complex chaotic behavior with two methods.
First,  we determine whether the collision dynamics is chaotic,  by using the existing binary test for chaos.  Finally,  we defined the constrained correlation embedding to quantify the dynamical complexity of the combined milling state. 
\section{\label{sec:Model}Swarm Model}
To make progress, we considered a class of well-known 
deterministic swarming models for both discrete  \cite{AlbiStability2014, Bernoff, Levine} and continuous \cite{Bernoff,  Minguzzi} systems consisting of mobile agents moving under the influence of self-propulsion, nonlinear damping, and pairwise interaction forces.  In the absence of interactions, each swarmer tends to a fixed speed, which balances its self-propulsion and damping, that is translationally invariant \cite{hindes2020unstable}.  A simple model 
that captures the basic physics is
\begin{align}
\ddot{\bold{r}}_{i}= \big[\alpha -\beta|\dot{\bold{r}}_{i}|^{2}\big]\dot{\bold{r}}_{i}-\lambda\sum_{j\neq i} \partial_{\bold{r}_{i}}U(|\bold{r}_{j}-\bold{r}_{i}|)
\label{eq:SwarmModel}
\end{align} 
where $\bold{r}_{i}$ is the position-vector for the $i$th agent in two spatial
dimensions, $\alpha$ is a self-propulsion constant, $\beta$ is a
nonlinear damping constant, and $\lambda$ is a coupling
constant \cite{Levine,Minguzzi,DOrsagna,Romero2012}. The total number
of swarming agents is $N$, and each agent has unit mass.  Beyond providing a
basis for theoretical insights, Eq.(\ref{eq:SwarmModel}) has been implemented
in mixed-reality experiments with several robotics platforms including
autonomous ground, surface, and aerial
vehicles\cite{4209425,Szwaykowska2016}.

An example interaction potential that we consider in detail is the Morse potential, 
\begin{align}
U(r)=Ce^{-r/l}-e^{-r}.
\label{eq:Morse}
\end{align} 
This is a common model for implicit interactions with local repulsion and
attraction length scales, scaled as $l$ and $1$, respectively\cite{DOrsagna,Bernoff}. 
In the following, we assume that two interacting swarms are subject to the same underlying physics,
Eqs.(\ref{eq:SwarmModel}-\ref{eq:Morse}),  but with different initial
conditions.  Each swarm will be
initially separated by a large distance compared to the interaction
length scales, $l$ and $1$. Therefore, the interaction force an agent feels
will be at early times effectively confined to their own swarm, given the
exponential decay with distance implied by Eq.(\ref{eq:Morse}). 
%\subsection{\label{sec:Two Colliding Swarms}Two Colliding Swarms}

For the swarm flocks that are nearly aligned upon collision,  the relevant
initial-condition parameter is the distance between the two flocks as they
approach $x\!=\!0$, regardless of the direction of their velocities. This
distance is often called the impact parameter,  $\Delta y$,  in classical mechanics\cite{Landau1976Mechanics},  and it represents the closest distance the two flocks would approach in the absence of interaction forces.  
Depending on the value of impact parameter and the coupling strength after collision,  the two
flocks typically translate as one,  scatter or form a combined mill.  

Based on the previous
numerical experiments for different values of $\Delta y$ and $\lambda$
\cite{CollidingSwarms2021},  we are particularly interested in collisions that
result in a swarm milling state.  Representative parameters for generating milling states  upon collision are: $l = 0.5, \alpha =1.0, \beta = 5.0,
\lambda = 0.1, \Delta y = 4.0$, which are assumed throughout this work.  See \cite{CollidingSwarms2021} for further details. Our aim is to vary the repulsive strength and observe the changes in the
dynamic behaviors of milling after the collision.  Having noticed that changing repulsion drastically changes the post-collision dynamics,  we then turned our attention to quantify the complexity of the dynamics so that we can try to deduce the dynamic consequences as we sweep parameters.  One useful technique is Karhunen-Loeve decomposition, which we address next. 
\section{\label{sec:Dim}Dimensionality Analysis: Karhunen-Loeve decomposition of swarms of N-agents}
Karhunen-Loeve (KL) decomposition is a procedure of mode expansion for
spatio-temporal processes that extracts the relevant degrees of freedom of the
dynamics of a large data set.  We measured the number of Karhunen-Loeve(KL) modes needed to capture most of the the dynamic variance.   It is an effective way to learn model reduction while describing the space in which the asymptotic dynamics resides.  Known as the method of snapshots \cite{Sirovich1987}, KL analysis correlates dynamics in time while averaging in space.  Note that KL mode decomposition essentially is Principle Component Analysis in the time domain \cite{PCA2016}.  In order to identify correlations,  we compute the covariance matrix,  where each entry is correlation of two snapshots. The modes are defined by the data and constitute a natural coordinate system that approximates from time-series data optimally in space with the $L_2$ norm.   From Eq.~\ref{eq:SwarmModel}, assume we have $N$ agents, which are sampled at $M$ time points.  Since both positions and velocities are changing with time, we define the sequence of data snapshots to
include both position and velocity by letting 
\begin{align}
 \bold{z}(t_j) = [\bold{r}_1(t_j),...,\bold{r}_N(t_j); \bold{v}_1(t_j),...,\bold{v}_N(t_j)].
\label{eq:PV}
\end{align} 
where $j\in\{1,2,...,M\}$ \cite{Rao1976KL}.  
We assume the field can be expanded in a split time-space sum of modes given
by
\begin{align}
\bold{z}(t_j) = \sum_{i = 1}^M{\alpha}_i(t_j)\bm{\phi}_i = \mathcal{A}_{j,i}\bm{\phi}_i.
\label{eq:Split}
\end{align}
The matrix $\mathcal{A}_{j,i}$ stores the elements of the amplitudes at time $t_j$ as rows.
That is, each row of the array is one mode. The vectors $\bm{\phi}_i$ are of the same
dimension as of $\bold{z}(t_i)$ filed at a given snapshot. 
We define the residual using $M$ modes over time as
\begin{align}
\mathcal{E} = \lVert \bold{z}(t_j) - \sum_{i = 1}^M{\alpha}_i(t_j)\bm{\phi}_i\rVert_{L_2}, \; j \in\{1,2,... M\}.
\label{eq:Res}
\end{align}
Minimizing $\mathcal{E}$ generates an eigenvalue problem for the
first temporal mode; i.e.,  we solve
\begin{align}
\mathcal{C}\bm{\alpha}_i = \lambda_i\bm{\alpha}_i ,
\label{eq:FirstTemp}
\end{align}
where $\mathcal{C}$ is an $M \times M$ covariance matrix.   The $ij^{th}$ element is given by
\begin{align}
[\mathcal{C}]_{ij} = \frac{<\bold{z}(t_i), \bold{z}(t_j)>}{N},
\label{eq:InnerProd1}
\end{align}
$< >$ denotes an inner product.  Since the covariance matrix is symmetric,   the eigenvalues are real,  non-negative. Then we have ordering such that 
\begin{align}
\lambda_1 \geq\lambda_2 \geq... \geq\lambda_M\geq 0.
\label{eq:Eigvals}
\end{align}
The corresponding eigenvectors, $\bm{\alpha}_i$, are the temporal modes. \\
To find the spatial modes, we make use of the fact that the modes from $\mathcal{C}$
are orthonormal; 
\begin{align}
\sum_{j = 1}^M{\alpha}_l(t_j){\alpha}_m(t_j)= \delta_{lm}. 
\label{eq:Ortho}
\end{align}
Finally,  we have 
\begin{align}
 \sum_{i = 1}^M\bold{z}(t_j){\alpha}_i(t_j) =  \sum_{j = 1}^M\sum_{i = 1}^M{\alpha}_i(t_j){\alpha}_k(t_j)\bm{\phi}_i = \bm{\phi}_k.
\label{eq:KL1}
\end{align}
We may now approximate the original spatial-temporal  field by examining the total variance, which may be computed from the eigenvalues.  Let
\begin{align}
\Lambda = \sum\lambda_i, 
\label{eq:KL2}
\end{align}
and normalize the eigenvalues by setting
\begin{align}
\bar{\lambda_i} = \frac{\lambda_i}{\Lambda}, i = 1,2,...M.
\label{eq:KL2}
\end{align}
\begin{figure}
\begin{center}
\includegraphics[width=9cm]{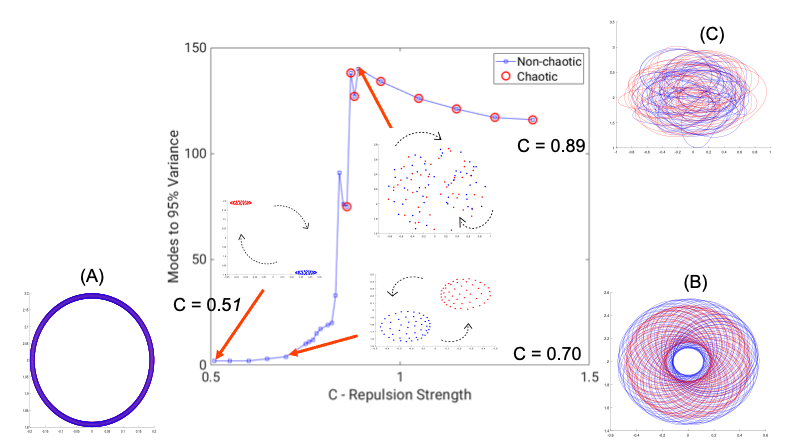}
\end{center} 
\caption{\label{fig:KL} The 0-1 chaos test result (red circles) plotted over KL mode decomposition (blue circles). The x
axis represents values of repulsive strength, C, and y axis is the number of KL modes needed to capture 95
percent of the total variance. The figures outside of the graph axis, (A)-(C), are the multi-agent attractor
time series projected onto the x-y plane corresponding to the time-series in Fig.~\ref{fig:TS}. The figures inside the axes are the corresponding fixed time snapshots showing the individual agents. This is to show that in real time, combined milling of two swarms are fragmented while it is not clear from the multi-agent attractor
time-series projected onto the x-y plane.}
\end{figure} 
We examine the cumulative sum of $\bar{\lambda_i}$ so that the total variance
exceeds a threshold, $T.$ Here  $T$ is arbitrarily chosen to be 95 percent.  After transients are removed,  in order to capture 95 percent of the full data,  for instance,  Fig.~\ref{fig:KL} indicates that 4 KL modes are needed for C= 0.51,  6 modes for C = 0.70 and about 140 modes for C = 0.89.    In other words,  one can see that when the post collision dynamics is periodic,  it only requires four modes to describe the effective limit cycle behavior.  As the repulsion strength is increased,  the dynamics lies on a torus, which may be thought of as two coupled oscillators.  However, the torus then breaks up,  as the repulsion is further increased and the resulting transition is to a chaotic behavior, corresponding to a substantial change in KL dimension,  from 4 to 140.  The rapid change in KL dimension signals a dramatic transition in dynamical complexity.   As we increase the repulsion strength,  the combined milling behavior becomes qualitatively more complex exponentially from collective oscillations.  Since the dynamics becomes quite complex and seemingly chaotic, we next turn to formally extracting whether or not is is technically chaotic for large repulsion.  

\section{\label{sec:Z1} Binary Test for  Chaos: 0 - 1 Test }
For its ease of use and vast applicability,  we employed the 0-1
test \cite{Gottwald04anew,  Zero1Test2009} to distinguish chaotic and non-chaotic parameter regions in our systems.  We input one-dimensional time series $\phi(n), n\leq
\bar{N}\in\mathbb{N}$ where $\bar{N}$ denotes the number of data points used.  and it produces two-dimensional Euclidean extension $P(n)$ and $Q(n)$. The main idea is to assess  whether the trajectories of the systems are bounded. 
 Let $\phi(n)$ be an observed time-series of the underlying system. \cite{Gottwald04anew} .  For an arbitrary $a > 0$ that is randomly chosen,  usually from $(0, \pi)$,  we define $P(n)$ and  $Q(n)$ for  each $a$ as
\begin{align}
P_a(n)= \sum_{j = 1}^n\phi(n)\cos(ja),\;\; Q_a(n)=\sum_{j = 1}^n\phi(n)\sin(ja).
\label{eq:PQ}
\end{align}
\begin{figure*}
\begin{center}
\includegraphics[width=5.925cm]{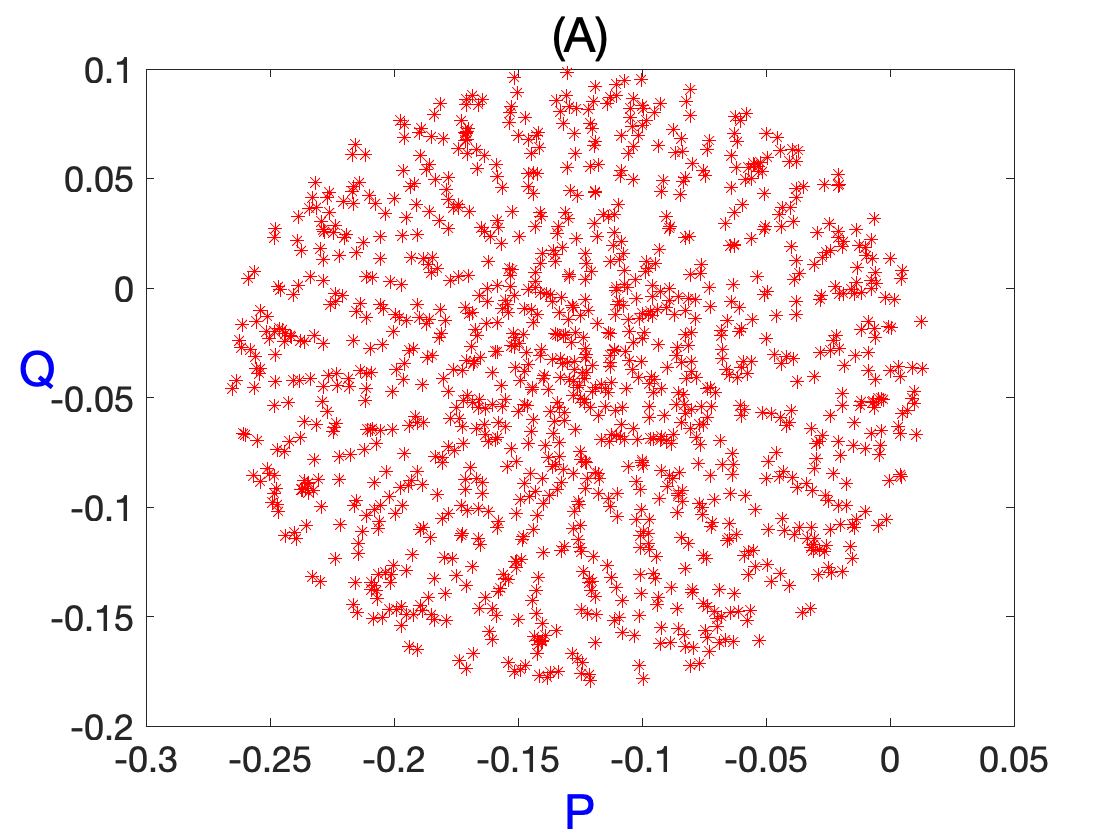}
\includegraphics[width=5.925cm]{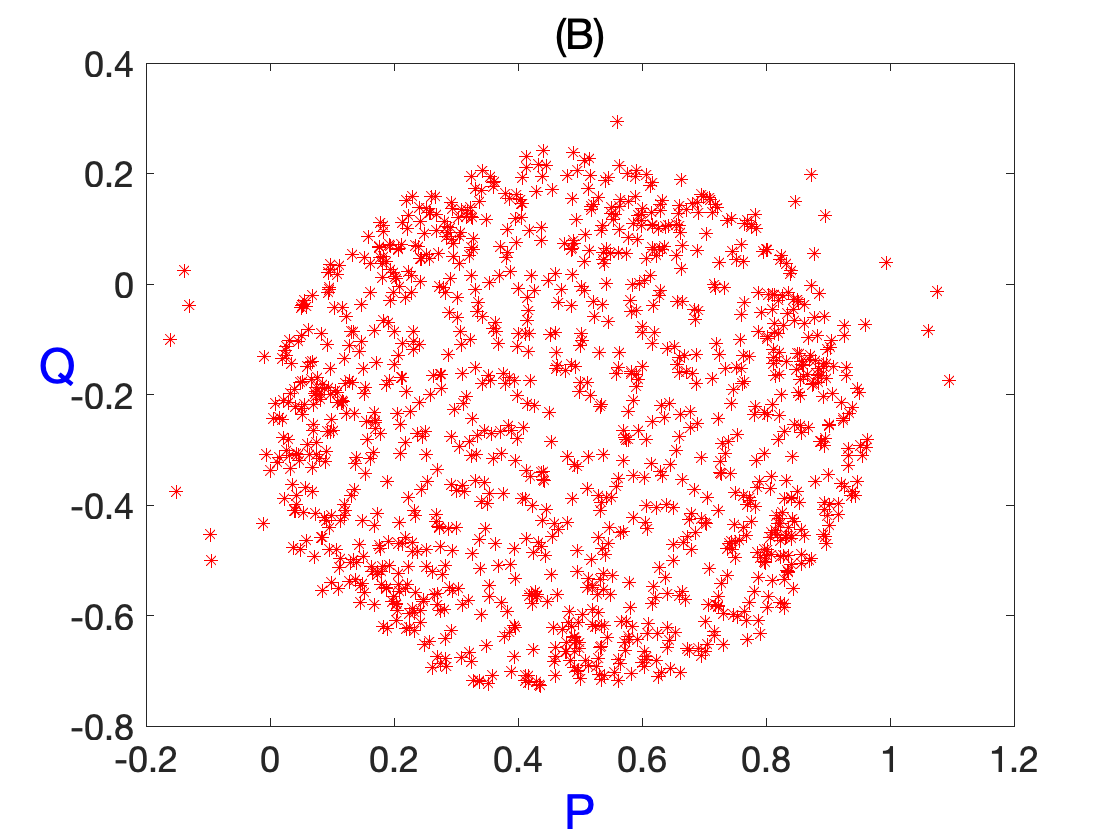}
\includegraphics[width=5.925cm]{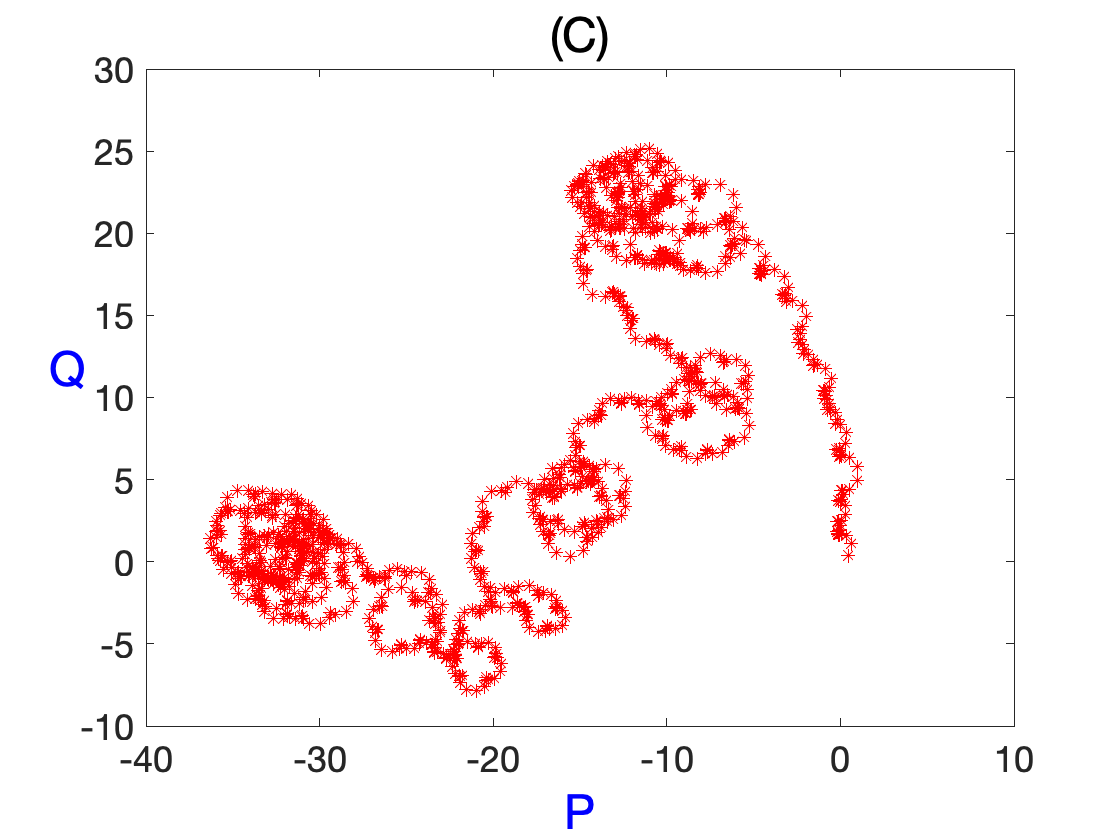}
\end{center}
\caption{\label{fig:PQ} P vs Q plot obtained  for  (A): C = 0.51,  (B): C = 0.70  and (C): C = 0.89.  P and Q are obtained in the process of 0 -1 test.  Underling system is regular if the PQ plot is bounded, and it is said to be chaotic if it is diffusive.  }
\end{figure*}
We then compute the mean square displacement defined by 
\begin{align}
\begin{split}
M_a(n) = & \lim_{\bar{N}\to\infty}\frac{1}{\bar{N}}\sum_{j = 1}^{\bar{N}}([P_a(j +n) - P_a(j)]^2\\ 
+ & [Q_a(j +n) - Q_a(j)]^2)], 
\end{split}
\label{eq:MSD1}
\end{align}
requiring $\bar{N}\gg n$.  The limit is guaranteed by calculating Eq.\ref{eq:MSD1} only for $n\leq n_{\text{cut}}$ where $\bar{N} \gg n_{\text{cut}}$, and $n_{\text{cut}} = \frac{\bar{N}}{10}$\cite{Zero1Test2009}. The test for chaos is based on the growth rate of the mean square displacement  as a function of $n$. 
We calculated the asymptotic growth rate of Eq.\ref{eq:MSD1}  defined as 
\begin{align}
K_a = \lim_{\bar{N}\to\infty}\frac{log(M_a(n))}{log(n)}. 
\label{eq:GrowthR}
\end{align}
Eq.\ref{eq:MSD1} is bounded if $\phi(n)$ is regular,  or it grows linearly in time if $\phi(n)$ is chaotic \cite{Gottwald_2009valid}.  Therefore,  the 0-1 test states that a value of $K \approx 0$ indicates regular dynamics, and $ K \approx  1 $ indicates chaotic dynamics.
For continuous systems such as differential equations,  one might encounter an inaccurate test  result  due to oversampling  \cite{Melbourne_2008, Zero1Test2009}.  We address this issue by incorporating two routine checks suggested in  \cite{MatlabMatthews}.  If the underlying dynamical system is regular, then its PQ plot is bounded; it is chaotic if the plot is diffusive.  Figure ~\ref{fig:PQ} identifies with the dynamics observation from the time-series in Fig.~\ref{fig:TS}.  Note that the 0 -1 test only distinguishes chaos from regular dynamics, and is unable to detect quasi- periodicity. 
 \begin{figure*}
\begin{center}
\includegraphics[width=5.925cm]{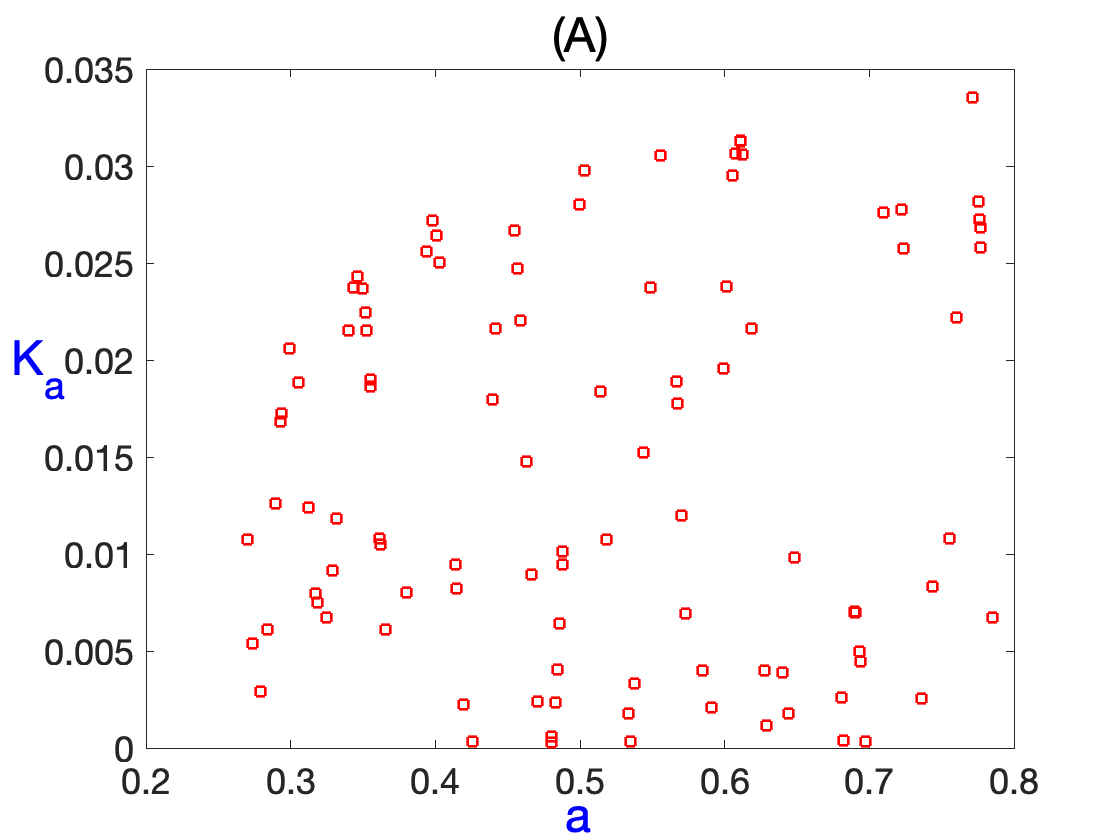}
\includegraphics[width=5.925cm]{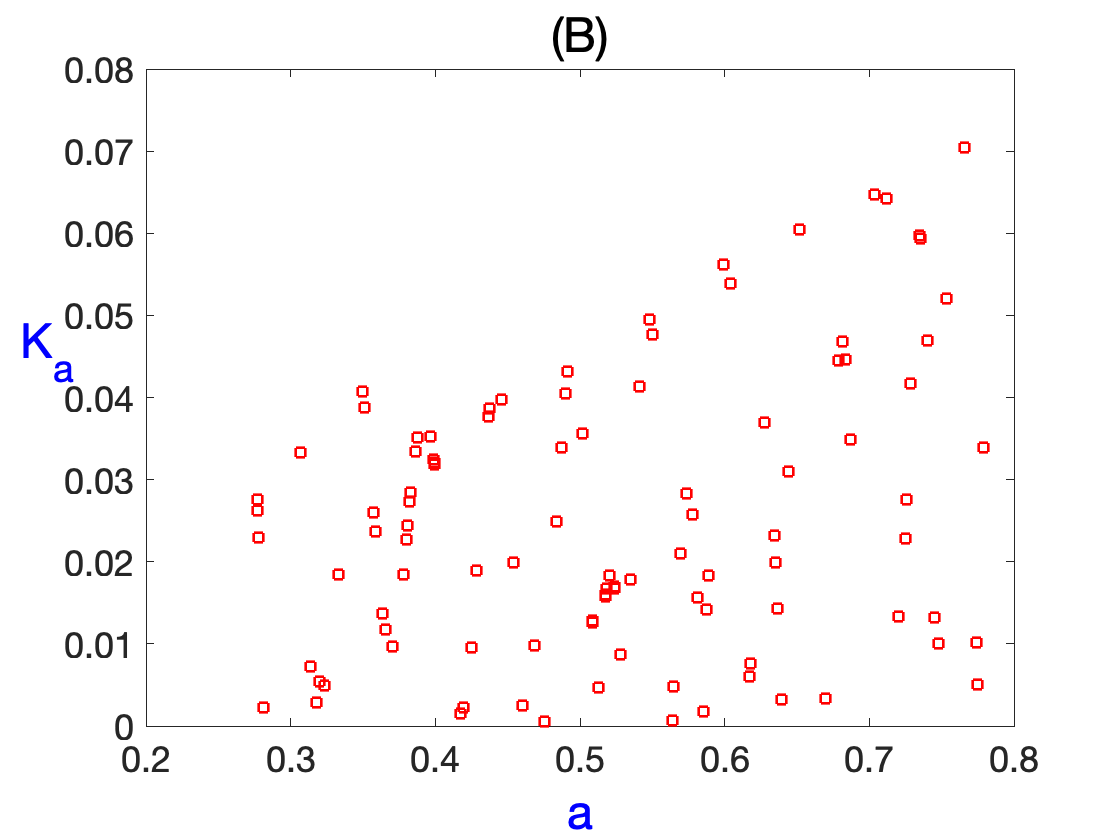}
\includegraphics[width=5.925cm]{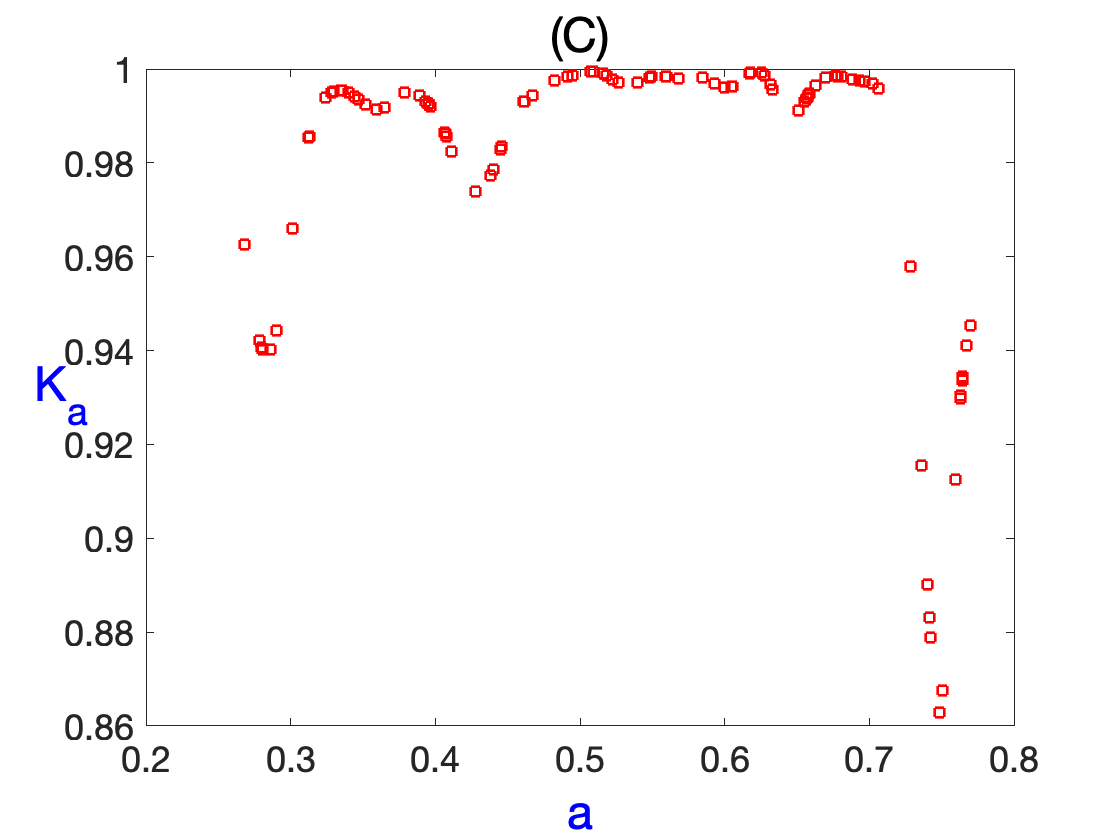}
\caption{\label{fig:Z1} 0 -1 test plots for  (A):C = 0.51,  (B):C = 0.70  and (C):C = 0.89. The result was computed for 100 randomly selected values of $a\in[\frac{\pi}{10}, \frac{2\pi}{10}]$ against corresponding value of $K_a$ for each such $a$.}
\end{center}
\end{figure*}
In our implementation of the 0 - 1 test,  we used 100 data points,  i.e.,  $N_a =100,$ from the time-series of y component of a randomly chosen agent.  It was collected from a post-collision milling state for various of repulsive strength $C$ \cite{CollidingSwarms2021},  excluding the transients.  We then calculated $K_a$ for each $a$ randomly chosen from $[\frac{\pi}{10}, \frac{2\pi}{10}]$ in accordance with an observation described in \cite{Gottwald_2008comment} to avoid oversampling (Fig.~\ref{fig:Z1}).  The indicator for chaotic behaviors, denoted as $K$, of our system for a given repulsive strength is then $K = $ median$(K_a)$.  We computed such $K$ as a function of varying repulsive strength $C$.  

\begin{figure}
\begin{center}
\includegraphics[width=9cm]{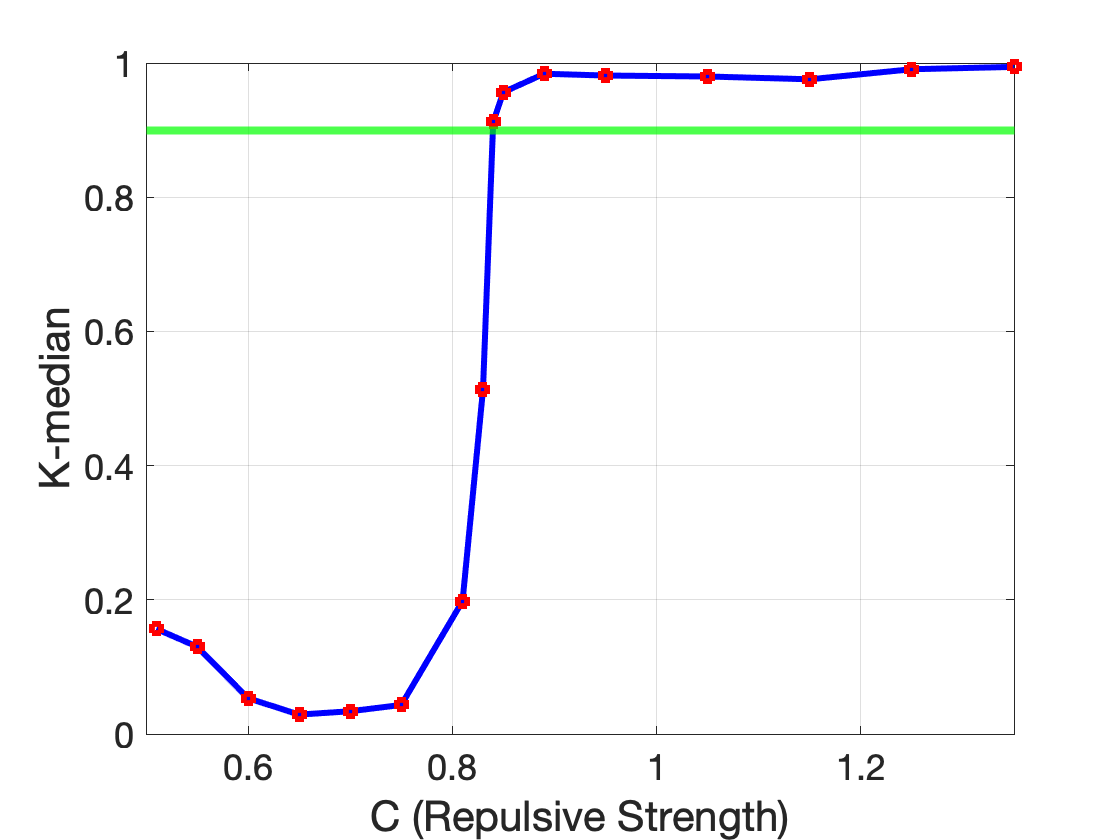}
\caption{\label{fig:Z1ALL} 0-1 test results obtained from plotting $K = $ median($K_a)$ for different values of the coupling strength C.  We interpret our result as chaotic for $K\geq 0.9$, indicated by green horizontal line.}
\end{center}
\end{figure}

Figure ~\ref{fig:Z1ALL} shows that the dynamics is regular for the repulsive strength $C$  that are in the range of 0.50  to 0. 81-83.  As the value of $C$ increases, the value of $K$ starts to tend to 1,  indicating chaos from onward around $C = 0.84$.  This roughly corresponds to the increasing trend of number of modes predicted by the KL decomposition depicted in Fig.~\ref{fig:KL}.  However, the 0-1 test does not provide any information about the specific properties or behaviors of the system beyond its chaotic or non-chaotic nature.  It cannot tell us anything about the attractors,  periodic orbits,  bifurcations, or other features that may be present in the system such as quasi periodicity.   For the current work,  we present our 0 -1 test result to illustrate what we have observed,   while other analytical techniques will be needed as a next step to fully understand and characterize the dynamics of the system under discussion. 
\section{\label{sec:CCE} The level of embedding of one swarm in another: Constrained Correlation Embedding}
In addition to detecting chaotic behaviors and categorizing dynamics state of milling after the collision,  we are interested to see the level of complexity of interaction between the two swarmsby looking at how one swarm embeds itself in another in a similar way as shown in the appendix of previous work \cite{SchwartzNATO2021},  the idea of which originated from fractal dimension.  Fractal dimension was developed as a way to quantify complexity of nature in relatively simple mathematical equations \cite{mandelbrot1983fractal}.  The estimation of the complexity becomes difficult in the case of the systems in which many factors of nonlinear interactions occur.  We use the notion of dimension in a locally constrained manner to evaluate and quantify the level of interaction between two swarms,  calling it``constrained correlation embedding.'' 

Without loss of generality,  we count the number of agents of swarm A in a local neighborhood of an agent of swarm B.  Let $N$ be a total number of agents in each Swarm.  For an agent $a_i \in A$ and an agent $b_j\in B$,  with the radius of a neighborhood  $\epsilon$,  we define
\begin{equation}
f(a_i, b_j,  \epsilon)= 
\begin{cases} 
     1, \; \text{ if }    ||a_{i}-b_{j}|| < &\epsilon , \\
     0, \; \text{ if } ||a_{i}-b_{j}|| \geq&\epsilon.
   \end{cases}
\label{eq:CCE1}   
\end{equation}
and 
\begin{align}
\gamma(\epsilon) = \sum_{i\in A}^N\sum_{j\in B}^Nf(a_i, b_j, \epsilon).
\label{eq:CCE2}   
\end{align}
$\gamma(\epsilon)$ is the total count of red agents within an $\epsilon$ neighborhood of each agent of swarm B.  We vary the radius of $\epsilon$ L times, and aim to see how the relative inclusion grows by plotting log of $\gamma(\epsilon)$versus log of $\epsilon$.  As the value of $\epsilon$ grows,  the number of agents from swarm A  in the $\epsilon$ ball centered on an agent from swarm B grows as the power law $\gamma(\epsilon) \propto \epsilon^d$.  The log-log plot behaves almost linearly until it saturates out for a large $\epsilon$.  The curve saturates at large $\epsilon$ because the $\epsilon$ balls will then engulf the whole swarm thus it cannot grow any further.  At extremely small values of $\epsilon$,  the only inclusion in each $\epsilon$ is just the k-th agent from swarm B itself.   For the range of $\epsilon$, let maximum distance and minimum distance of a given ith agent from swarm B with respect to all agents in swarm A be Max and Min, respectively.  Using the power law,  we divided the difference of Max and Min equally spaced logarithmically into 32\cite{HenryMathlabCode}, i.e., $L = 32$.

We computationally approximate $d$,  the slope of the line from the region that is most linear in the log-log plot,  and we defined it to be the contained correlation embedding,  much like the way correlation dimension is obtained\cite{Grassberger1983attractor} while it is done piece-wise \cite{strogatz:2000} but locally.   In order to evaluate the constrained correlation embedding, we sampled every 10th time steps,  totaling 1800 points of the time - series data of interacting swarms consisting of 50 agents each.  

\begin{figure}
\begin{center}
\includegraphics[width=8.5cm]{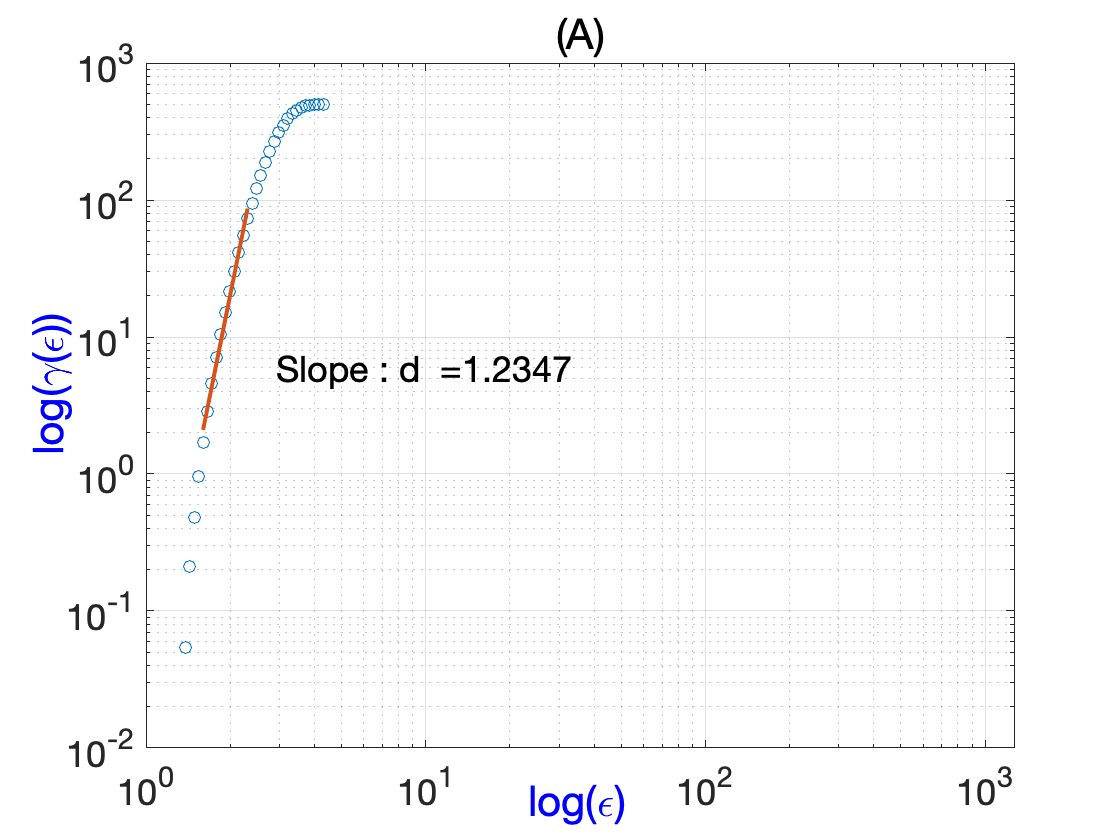}
\includegraphics[width=8.5cm]{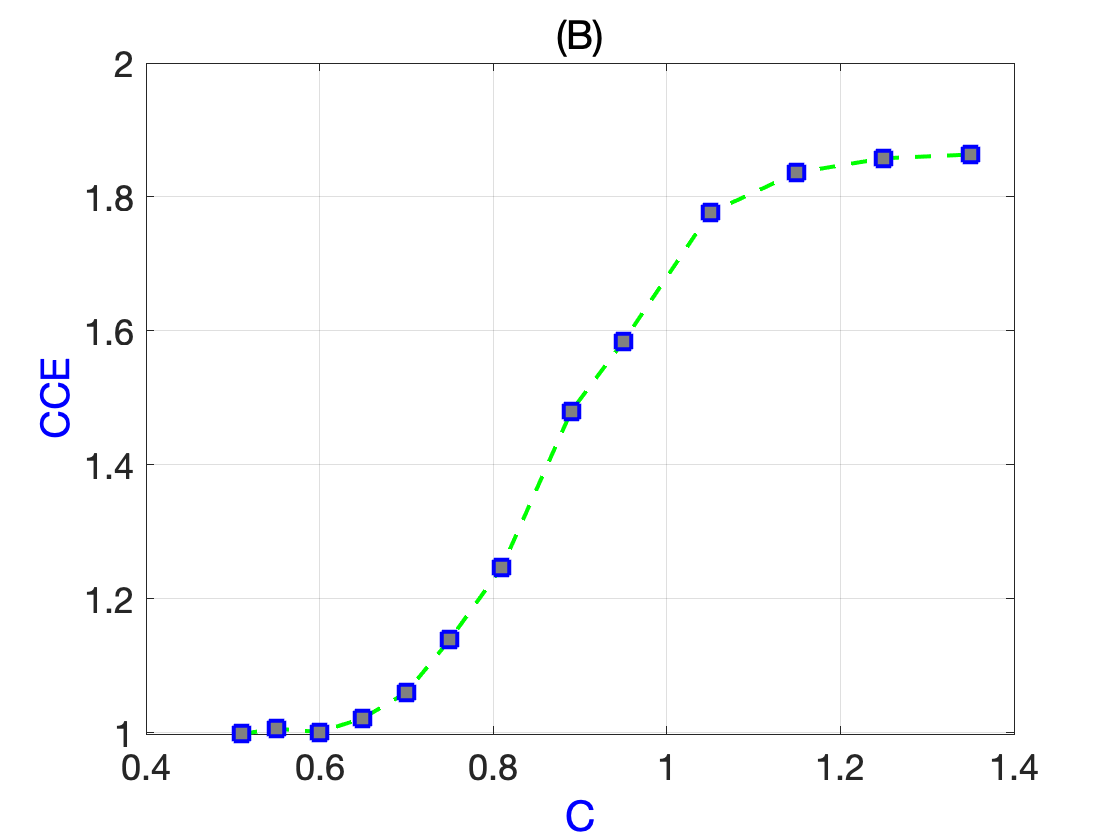}
\caption{\label{fig:CCE} (A) Constrained Correlation Embedding (CCE) for Repulsive strength C = 0.81, (B)  CCE as a function of repulsive strength C.}
\end{center}
\end{figure}

First,  our intuition tells us that a constrained correlation embedding should be between 1  and 2, since we are on the plane.  Figure ~\ref{fig:CCE}(A) is one representation of log- log plot at $C = 0.81$ and the slope $d$ reflects an embedding that is fractal.  The implication by CCE is that the agents of one swarm are embedded in another in a complicated way.  Furthermore, we see from Fig.~\ref{fig:CCE}(B), that the embedding grows linearly between  $C = 0.65$ and 1.05,   The steeper slopes indicate fast changes in the nature of the complexity.  On the other hand, we see that the complexity of milling behaviors of a combined swarms saturates beyond the value  $C = 1.05$ while the complexities up to $C = 1.05$ intensify for reach sampled value of $C$ from $C= 0.81$.  The same is true for the values of C below 0.65, which we interpreted as change of complexity that is more stable or minimal., consistent with our initial intuition. 

We have seen from  Fig.~\ref{fig:Z1ALL} that the milling state of two swarms
exhibit chaotic motion as the value of $C$ grows.  But the constrained
correlation embedding further illustrates increasingly complicated motions
within the chaotic region.  From Fig.~\ref{fig:CCE}(B), In references to the
supplementary material videos submitted,  we have detected and identified the milling behaviors that are consistent with the constrained correlation embedding tendency of Fig.~\ref{fig:CCE}(B): For $C\in[0.81, 0.95]$,  we say the combined swarm forms fragmented milling where the swarm mills but it still maintains the grouping of two.  Then for $C\in(0.95, 1.35]$, combined swarm forms (true) milling, where the combined swarm becomes more even and unified.  Notice that the milling at C = 0.95 still maintains grouping of 2 while  repulsive strength  C = 1.05 and C = 1.35, the milling behaviors changes only slightly,  reflecting the steepness of the slope.  (See supplementary material for videos of the dynamics
with varying repulsion.)
\section{\label{Conc}Conclusion, Discussion  and Future Works}
We have studied post collision dynamics of combined swarm milling states.  We
began our analysis by extracting the essential modal dynamics-information from
swarm collision data through a Karhunen-Loeve decomposition, then performed
the 0-1 test for detecting chaos; we further defined and used a constrained
correlation method to see the level of local complexity as a result of two
swarms colliding.  We have observed that a motion of periodic orbits
transitions to a motion on a torus before undergoing chaotic motion for a
given range of repulsion strength.  For a single swarm dynamics,  we have not
observed such transition to chaos using Eqs. \ref{eq:SwarmModel},\ref{eq:Morse}.  Furthermore,  our constrained correlation embedding method reveals that formation of the milling itself becomes more complex as a function of repulsive strength.  CCE enables us to categorize the milling as fragmented or true milling based on the range of repulsive strength;  where the simple binary test of chaos does not reveal the details of the complexity of the milling within chaotic region.  

One of the implications of the current result  is that
there is most likely a limitation to the theoretical mean-field analysis of the particular interacting swarms systems we studied.  The KL mode decomposition shows that the modes needed to capture 95 percent of the data variance increases drastically as the value of $C$ increases,
 and reveals a sharp transition to high dimensions.  It might be an indication
 that the bifurcation analysis at higher values of $C$ with the usual
 mean-field theory not be very meaningful due to spatial averaging.  We
 likely need to come up with a way that is more accommodating to the
 complexity and dynamical structural analysis of interacting swarms. 
 
Our future work includes the investigation of the sequence that leads to chaos,
as we have observed and have shown a possible Ruelle-Takens-Newhouse scenario
of chaos \cite {Gonzalez2023Scale} due to torus breakup, which is accompanied
by a large change in modal dimension as a function of $C$.  Moreover,
knowing the region of chaos in the swarm systems is a great advantage from a
control theoretical point of view \cite{Rosalie2016FromRP}.   We may be able
to stabilize unstable states of colliding swarm dynamics using various
control techniques,  such as those used in \cite{TriandafKLChaos1997},  which is based on analyzing the dynamics of the main coherent structure in the data represented by the highest energy  of KL mode. 

Lastly,  recall that from Eq.~\ref{eq:Split},  KL mode decomposition relies on space - time splitting hypothesis.   The concept of space-time splitting refers to the separation of variables into spatial and temporal components.  In some cases,  there  are dynamical systems in which the temporal behavior of the system cannot be separated from its spatial behavior.  In other words, the evolution of the system over time is inherently tied to the way in which it moves through space,  such as nonlinear waves \cite{Vadas2012Wave}, turbulence flow\cite{Manneville1991Turbulance} and chemical reactions \cite{Grasselli2015SpaceAT}.  
Overall,  many real-world dynamical systems do not exhibit the property of space-time splitting  \cite{Kirby2000geometric}.   Nonetheless,  in the current work,  we present a case where the rapid change in KL dimension correlates with a dramatic transition in dynamical complexity.  Combined with the constrained correlation embedding analysis,  we have enhanced insights to the complexity analysis of chaos. 

\section{\label{Sup}SUPPLEMENTARY MATERIAL}
The videos show the milling state of a combined swarms  consisting of
N=50 agents each.  The values of Repulsive strength C for five videos are: 0.51,  0.70, 0.89, 0.95, 1.05 and 1.35 respectively.  Other representative parameters for generating milling states after collision are fixed, as mentioned in Sec. \ref{sec:Model}. 
%a = 2:0; t = 1:75 The parameters for zero radius (delay is on
%all the time) are e = 0:0;cr = 0:05, and lr = 0:05 for a baseline,
%are shown in supplement S1 in Video1_eps_0p0.mp4.
%The parameters corresponding to Fig. 2 are e = 0:01;cr =
%0:01, and lr = 0:05 are shown in Supplement S2 in
%Video2_eps_0p01.mp4. The video shows that the attractor
%persists when repulsive forces are local and weak. Similar behavior
%is observed when N=150, which is used in Fig. 1 without
%repulsion; i.e., cr = 0. The parameters for corresponding
%to Fig. 3 are e = 0:25;cr = 0:05, and lr = 0:05, shown in
%Supplement S3 in Video3_eps_0p25.mp4.

\begin{acknowledgments}
\section{\label{sec:Acknowledgments}ACKNOWLEDGMENTS}
SK, JH and IBS were supported by the U.S. Naval Research Laboratory funding
(N0001419WX00055), the Office of Naval Research (N0001419WX01166) and
(N0001419WX01322), and the  Naval Innovative Science and Engineering. \\
\end{acknowledgments}
%\nocite{*}
\bibliography{ChaoticMillingSwarm, millingSwarmsOnSurfaces, DelayBistabilityPRE1, swarms, swarm_mendeley}
%millingSwarmsOnSurfaces,DelayBistabilityPRE1,swarms,swarm_mendeley}

\end{document}